\documentclass{article}[12pt]
\usepackage[a4paper, total={6in, 8in}]{geometry}
\usepackage[numbers]{natbib}
\usepackage{amsmath}
\usepackage{bookmark}
\usepackage{siunitx}

\usepackage{float,comment}
\usepackage{authblk}
\usepackage{graphicx}
\DeclareSIUnit \torr  { Torr }
\DeclareSIUnit \sccm  { sccm }

\title{Tracking the creation of single photon emitters in AlN by implantation and annealing}       
\author[1,2]{H. B. Ya\u{g}c{\i}}
\author[3,4]{E. Nieto Hernandez}
\author[1,2]{J. K. Cannon}
\author[1,2]{S. G. Bishop}
\author[3,4]{E. Corte}
\author[1,2]{J. P. Hadden}
\author[3,4]{P. Olivero}
\author[3,4]{J. Forneris}
\author[1,2]{A. J. Bennett}
\affil[1]{School of Engineering, Cardiff University, Queen's Buildings, 14-17 The Parade, CF24 3AA Cardiff, United Kingdom}	
\affil[2]{Translational Research Hub, Cardiff University, 1 Maindy Rd, CF24 4HQ Cardiff, United Kingdom}
\affil[3]{Dipartimento di Fisica e Centro Inter-Dipartimentale “NIS,” Universita di Torino, via Pietro Giuria 1, 10125 Torino, Italy}
\affil[4]{Istituto Nazionale di Fisica Nucleare (INFN), Sezione di Torino, via Pietro Giuria 1, 10125 Torino, Italy}
\date{}
\begin{document}
\maketitle
\vspace{-1cm}
\textit{Originally submitted to Optical Materials on 23 June 2024, accepted on 11 August 2024.}
\vspace{1cm}
\begin{abstract}
In this study, we inspect and analyze the effect of Al implantation into AlN by conducting confocal microscopy on the ion implanted regions, before and after implantation, followed by an annealing step. The independent effect of annealing is studied in an unimplanted control region, which showed that annealing alone does not produce new emitters. We observed that point-like emitters are created in the implanted regions after annealing by tracking individual locations in a lithographically patterned sample. The newly created quantum emitters show anti-bunching under ambient conditions and are spectrally similar to the previously discovered emitters in as-grown AlN.
\end{abstract}
\section{Introduction}\label{S:Intro}
\vspace{1mm}
\noindent Quantum technologies including quantum sensing, communications and computing have emerged as a potentially paradigm-changing sector, attracting strong scientific interest and  substantial investments \cite{McKinsey2024}. The major reason for this interest is the quantum advantage over classical information processing paradigms for specific and demanding computational problems. The advantage relies heavily on finding platforms hosting qubits that can be reliably prepared, manipulated and measured \cite{DiVincenzo2000}. One such platform is colour centres in wide-bandgap semiconductors. An extensively investigated system is the negatively charged nitrogen-vacancy centre in diamond (NV$^{-}$) \cite{Doherty2013}, which hosts an electron spin which can be manipulated using microwave fields and read out optically \cite{Clevenson2015}, even at room temperature. \\ \\
{\noindent}Other colour centres in semiconductors are being investigated for narrow transitions, high brightness, longer coherence lifetimes or ease of photonics integration; notable examples are SiV \cite{SiV_diamond} and SnV \cite{SnV_diamond} in diamond, divacancy centre \cite{DV_SiC} and V$_{\text{Si}}$ \cite{VSi_SiC} in silicon carbide (SiC), T- \cite{T_Si} and G- \cite{G_Si} centre in silicon and carbon-related defects \cite{HBN_odmr} in hexagonal boron nitride (hBN). Recently, quantum emitters have also been detected in III-nitrides, such as AlN \cite{Bishop2020AlN} and GaN \cite{GaN_Telecom}. Owing to the unique qualities of III-nitrides such as tunable bandgap and piezoelectricity, these emitters are promising candidates for electronically- or mechanically-modulated single photon sources in ambient conditions.\\ \\
{\noindent}AlN is hypothesised to host defect complexes that can be useful as potential qubits due to its large bandgap (\SI{6.02}{ e\volt}). Varley \textit{et al.} previously conducted an ab initio study on the vacancy complexes with group IVA and IVB elements, and reported that Ti-, Zr-vacancy centres can be hosted within the bandgap, with the ground state sufficiently apart from the valence band of AlN \cite{Varley_AlN}. Previously conducted experiments in Ti-implanted AlN found an electron spin resonance signal in implanted samples but could not confirm optical activity \cite{Aghdaei_Ti-AlN}. Aghdaei $et$ $al$. reported formation of broad peaks in Zr-implanted AlN \cite{Aghdaei_Zr-AlN}, but in a comprehensive study Senichev $et$ $al$. found that emission in Zr-implanted AlN is statistically not different than Kr-implanted AlN with similar ion damage \cite{Purdue_Zr-AlN}. In line with our previous observations in Al-implanted AlN \cite{Al_AlN}, these results point to a native origin for the created single-photon emitters (SPE). While these studies demonstrated SPE creation by statistically comparing different samples, the creation event of a single distinct emitter was not microscopically assessed. \\ \\
{\noindent}In this study, we rigorously inspect and analyse the effect of Al implantation into AlN by conducting confocal microscopy on the ion implanted regions, before and after implantation, followed by an annealing step. The independent effect of annealing is studied in an unimplanted control region, which showed that annealing alone does not produce new emitters. Through tracking individual locations in a lithographically patterned sample, we observe that point-like emitters are created in the implanted regions after annealing. The newly created quantum emitters show anti-bunching under ambient conditions and are spectrally similar to the previously discovered emitters in as-grown AlN.

\section{Materials and methods}\label{S:MatNMeth}
\vspace{1mm}
\noindent The studied samples consist of single crystal (0001) \SI{1}{\micro\metre} MOCVD-grown AlN epilayer on \SI{480}{\micro\metre} thick sapphire substrate, acquired from DOWA Electronics. SIMS measurements conducted by the manufacturer showed the main chemical impurities were H, C, O and Si, as is usual for samples grown by MOCVD. The epilayer in its as-grown form hosts single-photon emitters reported in our previous studies \cite{Bishop2020AlN}.

\subsection{Confocal microscopy}\label{SS:ConfocalHeader}
\vspace{1mm}
\noindent The sample was inspected through confocal microscopy with a dry objective with NA=0.9 under CW \SI{520}{\nano\metre} excitation. The emission was filtered through a \SI{550}{\nano\metre} long pass filter to suppress the  laser line, and a \SI{650}{\nano\metre} short pass filter to remove undesired background fluorescence from the sapphire substrate between 690-\SI{750}{\nano\metre} from the Cr$^{+3}$ impurity. The confocal microscopy scans were conducted with below-saturation excitation power (\SI{215}{\micro\watt} at the back of the  objective, compared to a typical saturation power of \SI{606}{\micro\watt}) to increase emitter-to-background contrast. The collected emission was guided to a fibre-coupled Excelitas silicon avalanche photodiode for photon counting measurements. As the excitation laser has a well-defined polarisation, the emitters whose absorption dipoles align with the excitation polarisation were detected preferentially against the emitters that did not. To counter this inherent bias, we conducted the confocal scans under linear polarisations of 0$^{\circ}$, 60$^{\circ}$ and 120$^{\circ}$ in the plane of the sample to maximise the probability of detecting changes in the scan maps as described in Cannon \textit{et al.} \cite{Joseph_Pol}. The individual intensity maps were then spatially aligned each other and the individual pixels were quadrature summed across three different polarisations in MATLAB. We note that the brightness of different emitters are not uniform. To show clearly the presence of less intense emitters in the scan maps presented in the manuscript, we have consistently rescaled the colour bar to the range between 1\% and 25\% of the raw image range. This means the brighter emitters are above the top end of the scale, and single pixel noise in the regions between emitters is less visible.\\ \\
{\noindent}For the Hanbury-Brown and Twiss (HBT) intensity interferometry, the filtered emission was split between two APDs using a multimode fibre beamsplitter. For each emitter, power dependent data was recorded to determine the best signal-to-noise ratio. Coincidence events from APD back-flash reflections were masked when fitting to the correlation functions. For the spectral measurements, the emission was coupled to an Andor Kymera 328i spectrometer with a grating of 150 l/mm and a silicon CCD camera through a polarisation-maintaining fibre. The emission spectrum was captured under maximum excitation power (\SI{3.7}{\milli\watt}) and laser polarisation aligned to the absorption dipole angle of the individual emitters. 
\subsection{Sample preparation}\label{SS:Markers}
\vspace{1mm}
{\noindent}The as-grown epilayer of AlN contains a random distribution of point-like emitters which were mapped before implantation to track the creation of new colour centres. The epilayer was patterned using photolithography, a hardmask and a Cl$_{2}$/BCl$_{3}$ dry etch. The hardmask was then removed and the sample cleaned in solvents. The sample was subjected to a \SI{100}{\watt} \SI{40}{\sccm} O\textsubscript{2} ashing step to remove any surface contaminants. Two marked regions of \SI{25}{\micro\metre} x \SI{25}{\micro\metre}, separated by \SI{1}{\milli\metre}, were dedicated to the zero fluence (control) and $10^{15}$ ions/cm$^2$ implanted area on the same \SI{5}{\milli\metre} x \SI{5}{\milli\metre} chip. The regions were subsequently inspected under scanning laser confocal microscopy.\\ \\
{\noindent}The sample was implanted with \SI{70}{\kilo  e\volt} Al ions with a multi-element ion source at the University of Torino. The implantation region was implanted with an ion beam masked through a \SI{400}{\micro\metre} x \SI{400}{\micro\metre} aperture for a fluence of 10$^{15}$ ions/cm$^2$. The vacancy distribution within the epilayer was simulated with the SRIM software package \cite{SRIM}, showing ion (vacancy) distribution peaking around \SI{87}{} (\SI{57}{}) \SI{}{\nano\metre} below the epilayer surface. The effect of initial ion implantation does not extend further than \SI{200}{\nano\metre} into the material.\\ \\
{\noindent}We conducted two annealing steps at \SI{600}{\degreeCelsius} and \SI{800}{\degreeCelsius} to thermally activate the implanted emitters. Both steps were conducted under nitrogen atmosphere and at atmospheric pressure with a JIPELEC JetFirst300 rapid thermal annealer in Cardiff University. The initial temperature ramp was set to \SI{1600}{\degreeCelsius\per\hour} and the sample stayed at the target temperature for 30 minutes, after which it was left to thermalise to room temperature. At all stages, the nitrogen flow was set to 300 sccm.
\section{Results}\label{S:ResultsHeader}
\subsection{Depth inspection on the native emitters} \label{SS:DepthProfResult}
\begin{figure}[h!]
	\centering
	\includegraphics[width=.6\textwidth]{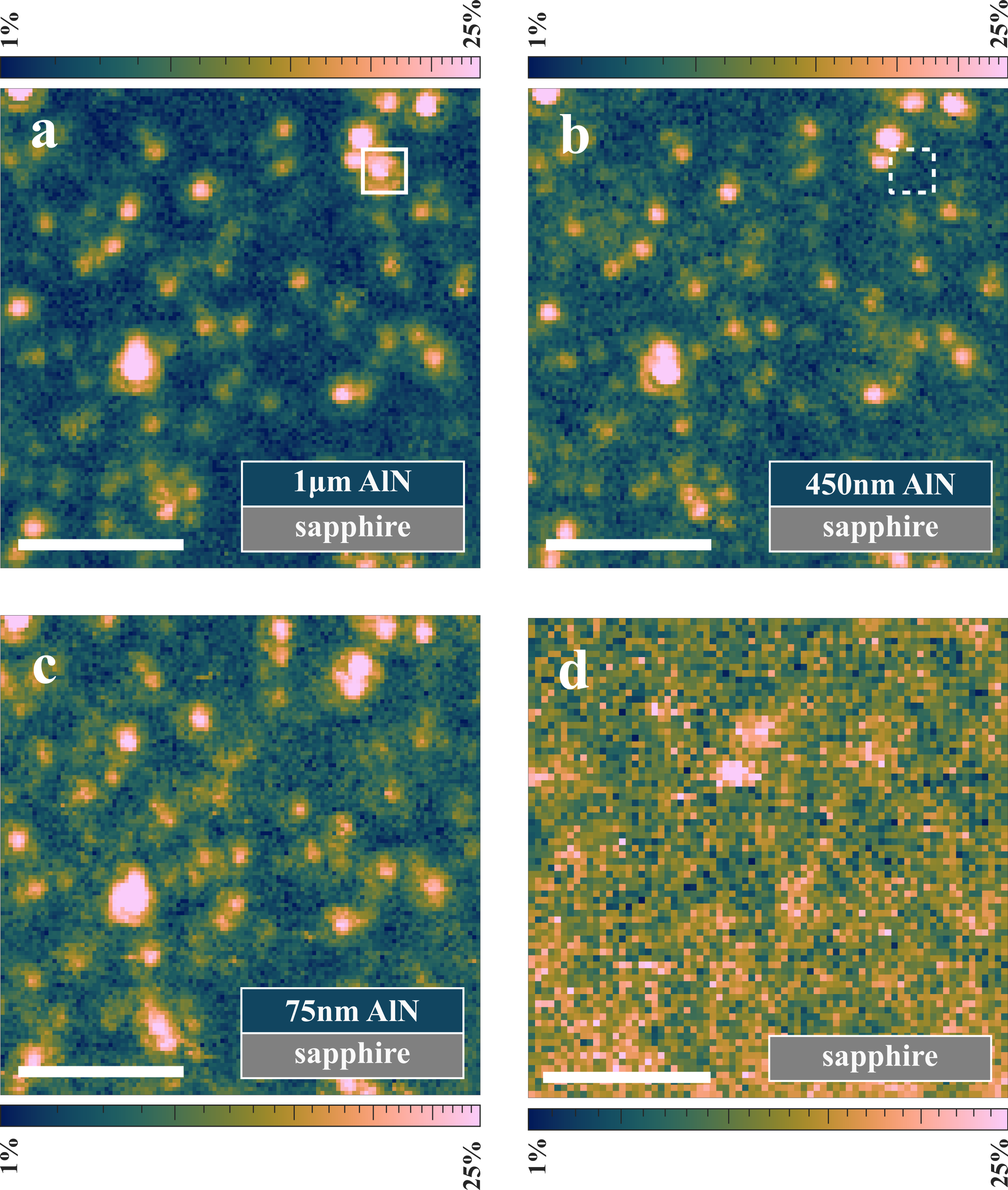}
	\caption{Investigating the depth of quantum emitters in the as-grown material as it is etched. Scale bar is \SI{5}{\micro \metre}.\textbf{a)} The epilayer at its initial thickness  of \SI{1}{\micro \metre}. The emitter which disappears in (b) is marked. \textbf{b)} The epilayer at a thickness of \SI{450}{\nano\metre}. \textbf{c)} The epilayer at a thickness of \SI{75}{\nano\metre}. \textbf{d)} The epilayer completely etched, with the substrate exposed.}
	\label{FIG:EnM}
\end{figure}

\vspace{1mm}
\noindent To interpret the impact of implantation and annealing in the epilayer, the depth distribution of the as-grown emitters was investigated. We took a direct approach and studied the density of emitters through repetitive material removal and confocal inspection steps. This enabled determining the depth of the emitters with accuracy determined by the etch, rather than by the Rayleigh range of the confocal mapping system. \\ \\
{\noindent}Following the lithographic marking, the sample was inspected under scanning confocal photoluminescence microscopy. The \SI{1}{\micro\metre} AlN epilayer showed an emitter density of \SI{0.22}{\micro\metre}$^{-2}$. 18 of the bright and isolated emitters, all showing anti-bunched emission statistics, were selected for tracking in the subsequent inspections. The AlN layer was then iteratively etched and scanned, effectively capturing the depth-dependent emitter distribution within the epilayer. Between each etching step and confocal inspection, the sample was cleaned with a 100 W oxygen plasma for 5 minutes to remove any remaining etch products from the surface. \\ \\
{\noindent}We observed no substantial change in the emitter distribution in the first few etch steps, as 93 \% of the epilayer is removed. Only one area (marked in Fig. \ref{FIG:EnM}a) changed in the first step. All emitters that were showing anti-bunched emission statistics (lack of coincidences between two detection events for small values of inter-path delay) were still observable at a layer thickness of \SI{75}{\nano\metre} (Fig. \ref{FIG:EnM}c). The epilayer was then removed at a final etch step, exposing the sapphire substrate. Subsequent inspection revealed that all emitters had been removed (Fig. \ref{FIG:EnM}d). This suggests that the native emitters are located at the first \SI{75}{\nano\metre} of AlN, close to the AlN/Sapphire surface. We note that this is the area in which the density of threading dislocations is greatest, and so emitter creation during MOCVD growth may be linked to this structural property as previously observed in GaN \cite{GaNMicrostructure}. Alternatively, the changing Fermi level near the sapphire interface might be populating the defect levels; the inverse of this effect was observed in NV\textsuperscript{-1} centres situated near Schottky barriers \cite{NVSchottky}. While the investigation of such effects is outside of the scope of this work, cathodoluminescence mapping of dislocations and electroluminescence experiments on as-grown defects can form the focal point of future studies.
\subsection{Effect of annealing in the control region}
\label{SS:IonImpControlAnneal}
\begin{figure}
	\centering
	\includegraphics[width=.6\textwidth]{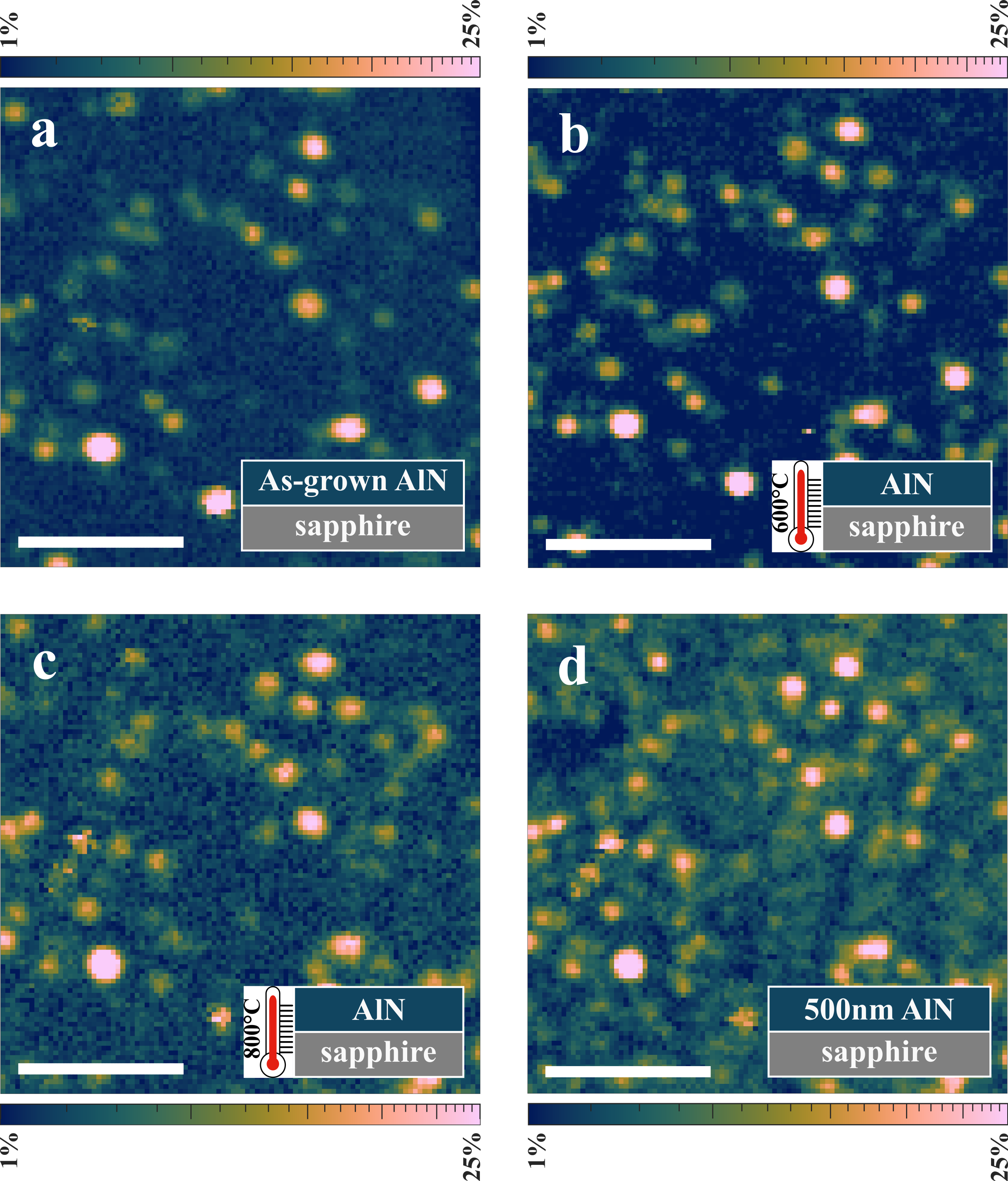}
	\caption{Effect of annealing on the as-grown (unimplanted) control area of the AlN chip. \textbf{a)} Control region before any thermal processing. \textbf{b)} Control region after annealing at \SI{600}{\degreeCelsius} for 30 minutes under N$_2$ atmosphere. \textbf{c)} Control region after annealing at \SI{800}{\degreeCelsius} for 30 minutes under N$_2$ atmosphere. \textbf{d)} Control region after removing \SI{500}{\nano\metre} of AlN epilayer. Scale bar is \SI{5}{\micro\metre}.}
	\label{FIG:Control}
\end{figure}

\vspace{1mm}
\noindent Following the depth profiling of the as-grown emitters, we studied the effect of annealing through step-wise confocal microscopy. Fig. \ref{FIG:Control}a shows the region prior to any annealing. The density of the native emitters was found to be \SI{0.19}{\micro\metre}$^{-2}$. Following the annealing at \SI{600}{\degreeCelsius} for 30 minutes (Fig. \ref{FIG:Control}b), we observed that the brightness of the emitters detected in the first inspection was non-uniformly modified. Overall, annealing at \SI{600}{\degreeCelsius} decreased the luminescence from the featureless "background" without removing or creating emitters.
This causes the emitters in Fig. \ref{FIG:Control}b to appear brighter compared to their pre-anneal scans (Fig. \ref{FIG:Control}a). Similarly, annealing at \SI{800}{\degreeCelsius} for 30 minutes (Fig. \ref{FIG:Control}c) did not create any new emitters in the control region, but instead lowered the brightness of the native emitters. Removal of the top \SI{500}{\nano\metre} of the epilayer (Fig. \ref{FIG:Control}d) did not result in removal of the native emitters, which again confirms the emitters are located in the region of the epilayer closest to the sapphire interface. As a result, we concluded that annealing at \SI{600}{\degreeCelsius} and \SI{800}{\degreeCelsius} for 30 minutes did not have a significant effect on the unimplanted control region.

\subsection{Effect of annealing on the implanted region} 
\label{SS:IonImp_Implanted} 
\begin{figure}[h!]
	\centering
	\includegraphics[width=.9\textwidth]{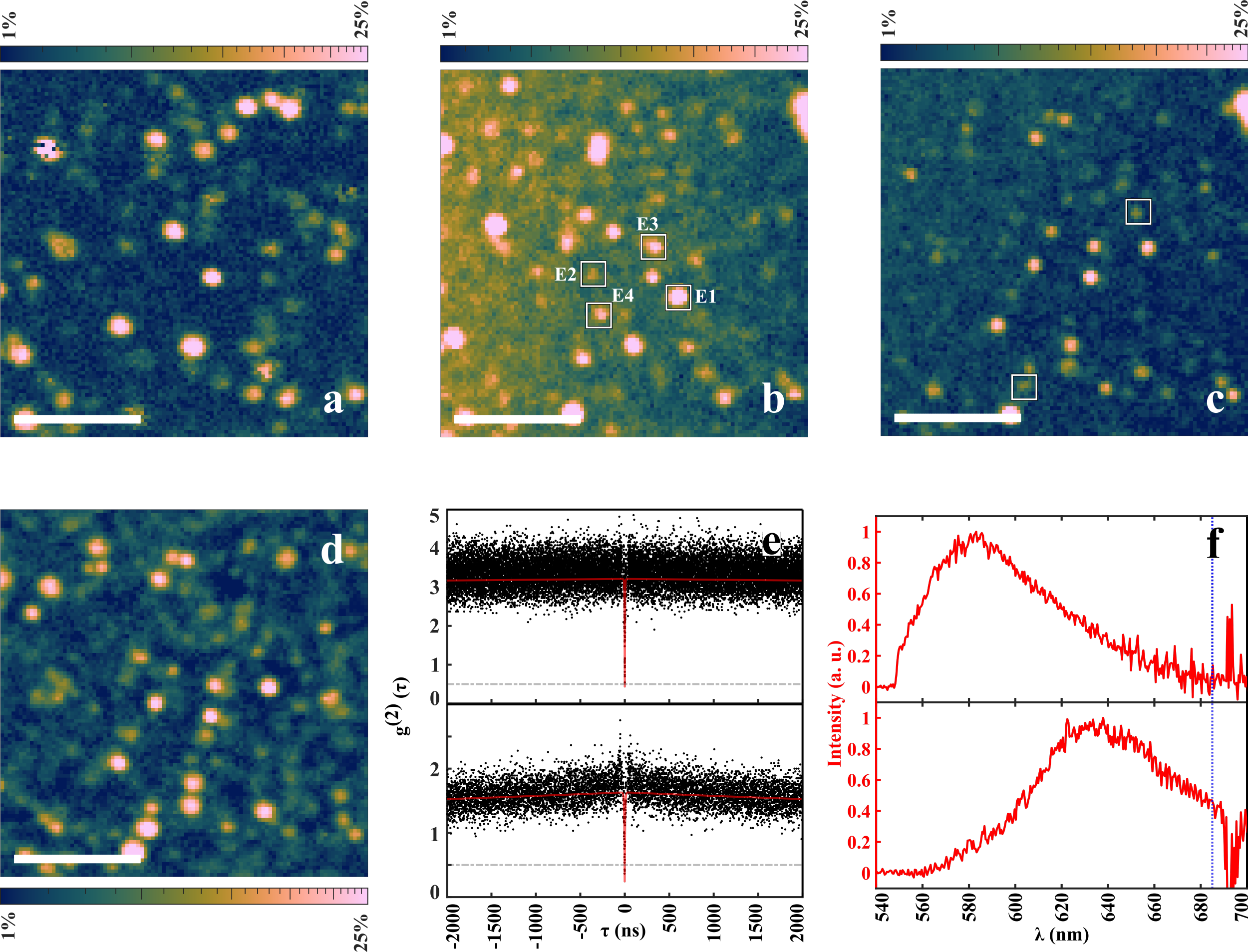}
	\caption{Confocal scan maps of the region dedicated to the fluence of 10$^{15}$ Al ions/cm$^2$. The scale bars are \SI{5}{\micro\metre}. \textbf{a)}: The region before implantation. \textbf{b)}: Implanted region after annealing at \SI{600}{\degreeCelsius} for  30 min under N$_2$ atmosphere. Some of the newly-created emitters are highlighted. \textbf{c)}: Implanted region after annealing at \SI{800}{\degreeCelsius} for 30 minutes under N$_2$ atmosphere. Some of the created emitters in (b) disappear (E1) and some new emitters appear (highlighted without label). \textbf{d)}: Implanted region after removing \SI{500}{\nano\metre} of AlN epilayer, effectively removing the region damaged by implantation. \textbf{e)} Hanbury-Brown and Twiss  interferograms of two emitters created by the implantation. E1 in (b) which disappears after \SI{800}{\degreeCelsius} anneal and E2 which does not disappear and survive the epilayer removal. \textbf{f)}: Spectra of E1 and E2 with the laser line removed with a \SI{550}{\nano\metre} long pass filter. The sharp features align with the Cr$^{+3}$ emission peak within the sapphire substrate. The blue dashed lines were included to represent the onset of this substrate luminescence.}
	\label{FIG:CreatedEmitters}
\end{figure}

\begin{figure}
	\centering
 \includegraphics[width=\textwidth]{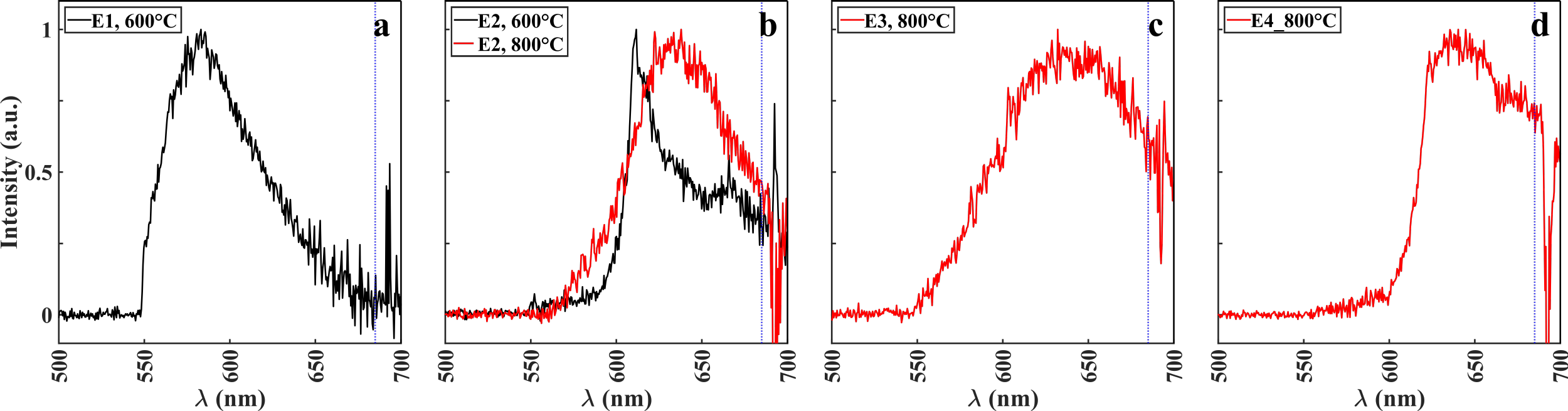}
\caption{Spectrograms for the newly-created emitters marked in Fig. \ref{FIG:CreatedEmitters}. The spectra obtained following the first (second) annealing step were plotted in black (red). Blue lines represent the onset of impurity luminescence from the sapphire substrate. All spectra were taken with a 550LP filter to filter the excitation. \textbf{a)} Spectra for E1 after 600 C annealing. \textbf{b)} Spectrum for E2, after 600 C (black) and after 800 C (red). \textbf{c)} Spectrum of E3 after 800 C annealing.\textbf{d)} Spectrum of E4 at 800 C.}
\label{FIG:NewEmitterSpectra}
\end{figure}

\vspace{1mm}
\noindent The region implanted with Al ions at  $10^{15}$ \SI{}{\centi\metre}\textsuperscript{-2} fluence can be seen after various stages of processing in Figure \ref{FIG:CreatedEmitters}. In panel \textbf{(a)}, the native emitters in the implanted region before implantation are illustrated. Post-implantation damage resulted in a high "background" luminescence dominating the whole area with no spatial structure, which displayed exposure-dependent bleaching. This is common in implantation studies, and is a transient result of the ion damage \cite{IonDamage}. \\ \\
{\noindent}After annealing at \SI{600}{\degreeCelsius} for 30 minutes under nitrogen atmosphere (Fig \ref{FIG:CreatedEmitters}b), new isolated emitters were detected in the scan maps. Two of these emitters (E1, E2) are highlighted in Fig. \ref{FIG:CreatedEmitters}b, which show antibunched photon statistics under ambient conditions. Both of the emitters displayed emission spectra similar to the previously-discovered emitters in AlN (Fig. \ref{FIG:NewEmitterSpectra}a-b) \cite{Bishop2020AlN}, with E2 having a distinct ZPL at \SI{611}{\nano\metre}. In total, 8 newly-created emitters showed up in the confocal scans over a \SI{25}{\micro\metre} x \SI{25}{\micro\metre} area, signifying a low yield in the creation of the emitters. Annealing at \SI{800}{\degreeCelsius} for 30 minutes produced some additional effects. Firstly, 4 emitters observed at the previous step ceased to fluoresce, including E1. Secondly, some of the emitters, such as E2, while still observable, displayed changed emission spectrum (Fig. \ref{FIG:NewEmitterSpectra}b, red). The second annealing step also caused diffusion of the previously-created emitters such as E4 and induced the creation of additional emitters (highlighted in Fig. \ref{FIG:CreatedEmitters}c). \\ \\
We did not observe a systematic difference between the spectral signatures of the as-grown emitters \cite{Bishop2020AlN} and the newly-created emitters (Figure \ref{FIG:NewEmitterSpectra}). In both cases, the variability in the signatures might be originating from non-homogeneous host environments \cite{CationVacancyZang}, occupation of different lattice sites  of identical defects \cite{BerhaneGaN}, or differences in defect species involved. The thermal stability of the as-grown emitters show that these emitters are in  energetically favoured configurations within the accessible parameter space of our annealer. In contrast, the relative instability of E1-like emitters suggest that these optically active emitters are formed when the constituents are nudged into a meta-stable formation energy local minimum during limited thermal treatment, only to morph into optically inactive, but possibly energetically favoured defect configurations during prolonged heat treatment \cite{G_Si}. \\ \\
To confirm quantum-mechanical emission coming from the newly-created emitters, Hanbury-Brown and Twiss interferometry was conducted on E1 and E2 with the setup detailed in Section \ref{SS:ConfocalHeader}. The correlations were normalised and fit with an empirical equation based on the state dynamics of a three-level model with a metastable shelving state:
\begin{equation}
g^{(2)}(\tau) = 1- a_1*exp^{-|\tau|/\lambda_1}+a_2*exp^{-|\tau|/\lambda_2} 
\end{equation} 
where $a_{1,2}$ are the amplitudes and $\lambda_{1,2}$ are the lifetimes of antibunching and bunching contributions, respectively. We observe that both E1 and E2 display anti-bunching below the conventional limit used to define a single quantum emitter ($g^{(2)}_{E1}(0)=0.375$, $g^{(2)}_{E2}(0)= 0.24$) with short lifetimes ($\lambda_1^{E1}=$ \SI{2.56}{\nano\s}, $\lambda_1^{E2}=$ \SI{4.46}{\nano\s}), albeit with significant bunching ($a_2 = 2.2, \lambda_2=$ \SI{10.5}{\micro\s} for E1; $a_2 = 0.63, \lambda_2=$ \SI{10.9}{\micro\s} for E2).  Such behaviour is consistent with our previous work \cite{Yanzhao_AlN}, where we report the existence of multiple long-lived metastable states in as-grown emitters in AlN on sapphire epilayers. While these metastable states are undesirable since they represent significant deviations from ideal two-level dynamics required for qubit applications and reduce the time-averaged intensity, their potentially spin-selective transitions to the ground level can be utilised to provide brightness contrast between different spin states in optically-detected magnetic resonance (ODMR) experiments. \\ \\
{\noindent}To further inspect the depth distribution of the created emitters, the epilayer was thinned by \SI{500}{\nano\metre} through dry etching. Surprisingly, the created emitters that still luminesce could still be observed after removing the upper half of the AlN epilayer, even though the initial damage caused by the implantation was confined in the first \SI{200}{\nano\metre}. This suggests that either the newly-created quantum-light emitters or its constituents were able to diffuse deep into the epilayer. As the implantation results in an excess Al content in the epilayer, we hypothesise the low-yield creation of these new emitters to complexes of interstitial Al, as Al$_{\text{i}}$ is harder to form compared to  V$_{\text{Al}}$ and V$_{\text{N}}$ but has similar \cite{Al_i_Migration} or larger \cite{AlPointDefectMigration} diffusivity  owing to a self-interstitial knock-out mechanism. Experimental studies exist for diffusion of Mg into AlN, which is thought to occur through a similar mechanism to Al$_{\text{i}}$ diffusion. Okumura \textit{et al.} in a recent study \cite{OkumuraAlN_1} compared depth profiles of different implanted dopants into a MOCVD-grown AlN sample and observed that Mg dopants were preferentially concentrated at AlN:Sapphire interface after annealing at elevated temperatures of $>$\SI{1300}{\degreeCelsius}. This may be rooted in dislocation-mediated diffusion promotion, which was previously observed in electrochemical etching of GaN in room temperature \cite{PorousGaNDiffusion}. The observed preference in the depth profile of the newly-created emitters in AlN points to an interstitial-related defect complex which is activated when in close proximity of the substrate. Further studies on the density of Al$_{\text{i}}$ within the epilayer, as well as DFT investigations focusing on Al$_{\text{i}}$-related defect complexes are needed for confirmation of this model.

\section{Conclusion}\label{S:ConclusionHeader}
\vspace{1mm}
In summary, we investigated the effects of ion implantation and annealing on Al-implanted and unimplanted aluminium nitride epilayers on sapphire. Through iterative confocal scans between each process step, we documented the creation of distinctive emitters within an implanted aluminium nitride epilayer, in contrast to the unimplanted regions which were unaffected by annealing. The spectral signatures of the new emitters were not systematically different than the native ones. We confirmed the created emitters displayed photon statistics consistent with single quantum emitters through Hanbury-Brown and Twiss  interferometry. The created emitters displayed large and long-lived bunching contributions, consistent with previous reports. Through thinning the implanted region, we reported the created emitters migrated deep within the epilayer (>\SI{500}{\nano\metre}), attributed to  diffusion of Al interstitials away from implantation depth. These results disclose new insight in the engineering of defect-based opto-electronic systems in a promising platform such as aluminium nitride, with appealing application in quantum communication and computing.
\section{Acknowledgements}\label{S:AckHeader}
\vspace{1mm}
\noindent This work was supported by the following research projects:\\
IR-HPHT, funded by the Italian Ministry of University and Research No. DM 737/2012 within the National Programme for Research (PNR); “Training on LASer fabrication and ION implantation of DEFects as quantum emitters” (LasIonDef) project funded by the European Research Council under the “Marie Skłodowska-Curie Innovative Training Networks” program; “Departments of Excellence” (No. L. 232/2016), funded by the Italian Ministry of Education, University and Research (MIUR); “Ex-post funding of research—2021” funded by Compagnia di San Paolo. The Project Nos. 20IND05 (QADeT) and 20FUN05 (SEQUME) and 20FUN02 (PoLight) leading to this publication have received funding from the EMPIR program co-financed by the Participating States and from the European Union's Horizon 2020 research and innovation program. P.O. gratefully acknowledges the support of “QuantDia” project funded by the Italian Ministry for Instruction, University and Research within the “FISR 2019” program. Cardiff University acknowledges financial support provided by UK's EPSRC via Grant No. EP/T017813/1 and EP/03982X/1 and the European Union's H2020 Marie Curie ITN project LasIonDef (GA No. 956387) and the Sêr Cymru National Research Network in Advanced Engineering and Materials.
\section{Data availability}\label{S:DataRepo}
\vspace{1mm}
\noindent The data that support the findings of this study are openly available in the Cardiff University Research Portal at \url{http://doi.org/10.17035/d.2024.0324890985}.

\bibliographystyle{model1-num-names}
\bibliography{article_bibliography}

\end{document}